\documentclass{kapproc} 
\let\footnote\savefootnote
\let\footnotetext\savefootnotetext
\let\footnoterule\savefootnoterule
\kluwerbib

\RequirePackage{graphicx}%
\RequirePackage{epsf}%
\input{psfig.sty}

\newcommand\lsim{\mathrel{\rlap{\lower4pt\hbox{\hskip1pt$\sim$}}
        \raise1pt\hbox{$<$}}}
\newcommand\gsim{\mathrel{\rlap{\lower4pt\hbox{\hskip1pt$\sim$}}
        \raise1pt\hbox{$>$}}}        

\begin{document}

\articletitle{Line versus Flux Statistics -- \\
Considerations for
the \\ Low Redshift Lyman-alpha Forest}

\author{Lam Hui}

\affil{Theoretical Astrophysics, Fermi National Accelerator Laboratory\\
Department of Physics, Columbia University}
\email{lhui@fnal.gov}

\chaptitlerunninghead{Line vs Flux Statistics}

\begin{abstract}
The flux/transmission power spectrum has become a popular 
statistical tool in studies of the high redshift ($z > 2$) 
Lyman-alpha forest. At low redshifts, where the forest has thinned out
into a series of well-isolated absorption lines, 
the motivation for flux statistics is less obvious. 
Here, we study the relative merits of flux versus line correlations,
and derive a simple condition under which one is favored over the other
on purely statistical grounds. Systematic errors probably play an important
role in this discussion, and they are outlined as well. 
\end{abstract}

\section{Introduction}  
\label{intro} 

Weinberg (this volume) has given a superb review of advances
in our understanding of the high redshift Lyman-alpha forest and its connection
to the cosmic web (Bond, Kofman \& Pogosyan 1996). Much recent work has focused on the 
the flux/transmission power spectrum, an 
approach pioneered by Croft et al. (1998) (see also Hui 1999). 
There are several different definitions in the literature. The one we adopt 
here is:
\begin{eqnarray}
\xi_f (\Delta v) = \langle \left[ {{f(v) - \bar f} \over \bar f} \right]
\left[ {{f(v + \Delta v) - \bar f} \over \bar f} \right] \rangle \\ \nonumber
P_f (k) = \int \xi_f (\Delta v) e^{-ik \Delta v} {d \Delta v}
\end{eqnarray}
where $\xi_f$ is the two-point flux correlation ($\Delta v$
specifies the lag in velocity), and
$P_f$ is its Fourier-transform, the flux power spectrum. 
Here $f$ is simply the transmission $f = e^{-\tau}$ where $\tau$ is the
Lyman-alpha optical depth. The symbol $\bar f$ denotes the mean transmission.
Finally, $k$ is the wave-number in units of inverse velocity. 

The flux-statistics above, which treats the transmission
fluctuations on a pixel-by-pixel basis, 
is motivated by a physical picture in which
the forest arises from continuous fluctuations in the intergalactic medium,
rather than discrete, well-isolated clouds (Bi, Boerner \& Chu 1992, Cen et al. 1994;
for additional ref., see 
Hui et al. 1997 and ref. therein). 
A second class of statistics, which has a longer history,
treats the transmission fluctuations on a line-by-line basis. 
The counting of absorption lines in terms of their properties, 
such as the column density distribution, falls into this category.
The analog of the {\it flux} two-point correlation, or power spectrum,
is the {\it line} correlation or power spectrum, defined as:
\begin{eqnarray}
\xi_{n} (\Delta v) = \langle \left[ {{n(v) - \bar n} \over \bar n} \right]
\left[ {{n(v + \Delta v) - \bar n} \over \bar n} \right] \rangle \\ \nonumber
P_{n} (k) = \int \xi_n (\Delta v) e^{-ik \Delta v} {d\Delta v}
\end{eqnarray}
where $n(v)$ is the number density of lines, and $\bar n$ is its mean. 
Implicit in this definition is that one studies the correlation
of absorption lines within some range of column density or equivalent width, or
above some threshold. 

The respective motivations for line and flux statistics are probably
both valid, depending on circumstances. For the low column density
forest which probably arises from smooth fluctuations, 
flux statistics seems reasonable. For the higher column density
systems, which likely arise from well-isolated galactic or 
pre-galactic halos, line statistics seems to provide a good characterization.
{\bf The aim of this short note is to ask a purely statistical question,
irrespective of the underlying physical picture: which kind of statistics
can one measure with more precision?}

\section{Statistical Error Analysis for the Flux vs. Line Power Spectrum}
\label{stat}

The statistical error can be worked out for both the two-point correlation
function and the power spectrum. The result is somewhat simpler to state
in Fourier space, and so we will focus on the power spectrum. 
The Fourier space description has the additional advantage that
the powers in separate wave-bands are uncorrelated, provided 
that the fluctuations are Gaussian random. The latter is a crucial
assumption in our discussion below -- the fluctuations in flux or
number density of lines are almost certainly not exactly Gaussian random.
However, because correlations seen in the forest are often quite weak,
Gaussianity is not a bad approximation; at least, it provides us
a way to gauge the relative importance of shot-noise and the correlation
signal, as we will see. By the central limit theorem, lower resolution
data also tend to be more Gaussian random. 

The statistical dispersion in the measured flux power spectrum is given by:
\begin{eqnarray}
\label{error1}
\langle \delta P_f (k)^2 \rangle^{1/2}
= {1\over \sqrt{N_k}} (P_f (k) + N_f^{-1})
\end{eqnarray}
where $N_k$ is the number of Fourier modes in the waveband of interest
(which is centered at $k$) i.e. if the waveband has a width of 
$\Delta k$, $N_k = \Delta k/(2\pi/L)$ where $L$ is the length of 
the quasar absorption spectrum (if one has more than one line of sight, 
one adds the error in quadrature in the usual way).

The quantity ${N_f}$ (not to be confused with $N_k$) gives us a measure
of the signal-to-noise of the data: the smaller $N_f$ is, the larger
the shot-noise. To be precise, 
\begin{eqnarray}
\label{shotnoise}
N_f^{-1} = {dv \over {\cal N}} \sum_i {{\rm var}(i) \over 
{\bar N_Q (i)}^2} \sim {dv \over \bar f} (N/S)^2
\end{eqnarray}
where $dv$ is the velocity width of each pixel, ${\cal N}$ is the
number of pixels, ${\rm var} (i)$ is the variance of counts
in pixel $i$, and $\bar N_Q (i)$ is the mean quasar photon count in pixel
$i$ (e.g. for a flat continuum, $\bar N_Q$ would be independent of $i$).
A useful approximation (accurate to
within a factor of two or so) to the shot-noise ${N_f}^{-1}$ is given by
$(dv/\bar f) (N/S)^2$ where $\bar f$ is the mean transmission as before,
and $N/S$ is the average noise-to-signal ratio at the level of the continuum.

Eq. (\ref{error1}) is derived in Hui, Burles et al. (2001). 
Its intuitive meaning is quite apparent if one writes down
the fractional error:
\begin{eqnarray}
\label{fracPf}
{\langle \delta P_f (k)^2 \rangle^{1/2} \over  P_f(k)}
= {1\over \sqrt{N_k}} (1 + [N_f P_f(k)]^{-1})
\end{eqnarray}
One can see that {\bf 1.} the longer the spectrum is, the larger the
number of modes $N_k$, and therefore the smaller the fractional error;
{\bf 2.} the larger the intrinsic signal (i.e. $P_f (k)$), the smaller
the fractional error; {\bf 3.} the more noisy the spectrum is, the
larger ${N_f}^{-1}$ is, and therefore the larger the error. 

{\bf How about the statistical error for the line power spectrum?}

The expression is very similar. The fractional error
for the line power spectrum is:
\begin{eqnarray}
\label{fracPn}
{\langle \delta P_n (k)^2 \rangle^{1/2} \over  P_n(k)}
= {1\over \sqrt{N_k}} (1 + [\bar n P_n(k)]^{-1})
\end{eqnarray}
where $P_n$ is the line power spectrum, $N_k$ is the same number
of modes in the waveband centered at $k$, and
$\bar n$ is the number density of lines. 
The intuitive meaning of this expression is also quite clear: 
the smaller the number density of lines, the larger the fractional error.
The only difference between eq. (\ref{fracPf}) and (\ref{fracPn})
is that ${N_f}^{-1}$ has been replaced by ${\bar n}^{-1}$.
In other words, shot-noise from photon-counts is replaced by shot-noise
from the finite number of absorption lines. 

Before we draw conclusions from these two expressions, we should note
that our results for the statistical error assume
the quadratic estimator for the respective power spectrum is of a
particular form (known in the large scale structure literature as 
$(DD - 2DR + RR)/RR$; Landy \& Szalay 1993); 
other forms generally lead to larger errors.
We refer the reader to the discussion in Hui et al. (2001) for details.
The discussion there focused on the flux statistics, but very similar
reasoning applies to line statistics as well. 

\section{Discussion}
\label{discuss}

Eq. (\ref{fracPf}) and (\ref{fracPn}) in the last section give
the respective fractional error in flux power spectrum and line power spectrum.
From the two expressions, it is plain to see that
the flux power spectrum can be measured with a higher
statistical precision than the line power spectrum {\it if}
\begin{eqnarray}
\label{Nf}
N_f > \bar n P_n / P_f
\end{eqnarray}
where $N_f \sim (S/N)^2 \bar f / dv$ (eq. [\ref{shotnoise}]) is
roughly the typical signal-to-noise-squared per km/s of the quasar spectrum,
$\bar n$ is the number density of lines, $P_n$ is the line power spectrum,
and $P_f$ is the flux power spectrum. 

At $z \sim 3$, all quantities on the right hand side have been measured,
so we can derive the condition on the $S/N$ above which the flux power spectrum
can be measured with greater precision. 
The result depends of course on the scale of interest. Let us pick a typical
scale of around $k \sim 0.01$ s/km (or velocity separation of about
300 km/s). At this scale, $P_n / P_f$ is about $100$, depending
on the column density of the absorption lines (a lower column density
cut of $\sim 10^{14} {\,\rm cm^{-2}}$; including more low
column density lines would decrease this ratio) (see Cristiani et al. 1997
and McDonald et al. 2000), while $\bar n \sim 2 \times 10^{-3} ({\,\rm km/s})^{-1}$
(Kim et al. 2002). 
Finally, $\bar f \sim 0.65$. 
Hence, the requirement for favoring flux over line power spectrum is:
\begin{eqnarray}
\label{SN}
(S/N)^2 /dv \, \gsim \, 0.3 ({\,\rm km/s})^{-1}
\end{eqnarray}
One can see that this is not a very stringent requirement on the signal-to-noise
at all. For high quality Keck spectra, signal-to-noise of several tens
per resolution element ($dv \sim 10$ km/s) is quite typical, and so
$(S/N)^2 / dv \gg 0.3 ({\,\rm km/s})^{-1}$. For noisy, low-resolution spectra such as
those obtained from the Sloan Digital Sky Survey, 
$(S/N)^2 /dv \sim 10^{-2} - 1$, it looks as though the line power spectrum
might be favored, but one must keep in mind that for low-resolution data,
both $\bar n$ and $P_n$ are much reduced, and the requirement on 
$(S/N)^2 /dv$ can be relaxed by as much as a factor of $100$. 

The situation at lower redshifts $z < 2$ is more uncertain. This is because
no measurements have been made of the flux power spectrum at low redshifts,
although much is known abut the absorption-line number density and clustering
(e.g. Weymann et al. 1998, 
Impey 1999, Penton et al. 2000, Dave \& Tripp 2001, Chen et al. 2001, Bechtold et al. 2002). 
Both $\bar n$ and $P_f$ drop as one goes to lower redshifts, although $P_n$ tends
to increase (this statement is cut-off dependent; we assume here a fixed column-density
or equivalent-width threshold). One possibility is to assume that
$\bar n P_n/P_f$ stays roughly constant, in which case eq. (\ref{SN}) remains
a valid requirement on the signal-to-noise of the data. Instruments on-board HST
frequently yield spectra that satisfy this requirement. 
It must be emphasized, however, $P_f$ has yet to be measured at low redshifts, 
and, if measured, one must go back to the expression in 
eq. (\ref{Nf}) to draw the appropriate conclusion.

To end our discussion, it is important to underscore the fact that our discussions
so far focus entirely on the issue of statistical error. Systematic errors could
make a significant difference to the conclusion one draws, as emphasized by several
members of the audience. Two sources of systematic errors were brought up.
One is that the efficiency of the spectrograph or detector 
might not be sufficiently well-characterized to allow an accurate flux correlation measurement.
However, if the efficiency has small-scale fluctuations that are not well-understood,
neither should one trust the absorption-line measurements. 
Second, spurious power introduced by the continuum might be more of an issue
for the flux correlation than for the line correlation. This is certainly
a potential worry. One should keep in mind, however, that continuum-fitting
is in fact easier at low redshifts than at high redshifts, because of the thinning
out of the forest (although continuum-fitting is actually not recommended as part
of the data reduction;
see Hui et al. 2001). The important question is: what is the scale below which
the forest fluctuation dominates over the continuum fluctuation (recall that
the continuum is smooth while the forest has lots of small scale structure)? 
At $z \sim 3$, this scale is about $k \sim 0.001$ s/km (or velocity separation of
about a few thousand km/s).
As one goes to lower redshifts, the forest flux power $P_f$ drops, and so 
this scale must move to a smaller value (or higher $k$). The issue is whether this scale is still
sufficiently large to be interesting. At the very least, the author hopes that
this short note will provide a stimulus to measure the flux power spectrum from
low redshift quasar spectra. Measurements from actual data are certainly far
more useful than speculations from a theorist. 



Thanks are due to the organizers of the IGM conference, especially
Mary Putnam and Jessica Rosenberg, for gently and patiently
urging the author to write up
his talk, and to Todd Tripp for useful discussions. The interest expressed by 
Chris Impey in the issues discussed here has also
provided an important motivation. This short paper covers the second half of
the conference presentation. For the first half on
the galaxy-IGM connection at $z \sim 3$, see Hui \& Sheth (2002, in preparation); for
related observational results, see Adelberger et al. (2002). 
Support for this work is provided by an Outstanding Junior Investigator
Award from the DOE, an AST-0098437 grant from the NSF, and by
the DOE at Fermilab, and NASA grant NAG5-10842.

\begin{chapthebibliography}{1}  

Adelberger, K. L., Steidel, C. C., Shapley, A. E., Pettini, M. 2002, astro-ph 0210314

Bechtold, J., Dobrzycki, A., Wilden, B., Morita, M., Scott, J., Dobrzycka, D., 
Tran, K., Aldcroft, T. L. 2002, ApJS, 140, 143

Bi, H. G., Boerner, G. \& Chu, Y. 1992, A \& A, 266, 1       

Bond, J. R., Kofman, L. \& Pogosyan, D. 1996, Nature, 380, 603

Cen, R., Miralda-Escude, J., Ostriker, J. P. \& Rauch, M. 1994, ApJL, 437

Chen, H. W., Lanzetta, K. M., Webb, J. K. \& Barcons, X. 2001, ApJ, 559, 654

Croft, R. A. C., Weinberg, D. H., Katz, N., Hernquist, L. 1998, ApJ, 495, 44

Cristiani, S., D'Odorico, S., D'Odorico, V., Fontana, A., Giallongo, E. \& Savaglio, S.
1997, MNRAS, 285, 209
 
Dave, R. \& Tripp, T. M. 2001, ApJ, 553, 528

Hui, L., Gnedin, N. Y., Zhang, Y. 1997, ApJ, 486, 599

Hui, L. 1999, ApJ, 516, 519 [astro-ph 9807068]
                                          
Hui, L., Burles, S., Seljak, U., Rutledge, R. E., Magnier, E., Tytler, D. 2001, ApJ, 552, 15

Impey, C. D., Petry, C. E. \& Flint, K. P. 1999, ApJ, 524, 536

Kim, T. S., Carswell, R. F., Cristiani, S., D'Odorico, S. \& Giallongo, E. 2002, MNRAS, 335, 555

Landy, S. D. \& Szalay, A. S. 1993, ApJ, 412, 64

McDonald, P., Miralda-Escude, J., Rauch, M., Sargent, W. L. W., Barlow, T., 
Cen, R. \& Ostriker, J. P. 2000, ApJ, 543, 1

Penton, S. V., Shull, J. M. \& Stocke, J. T. 2000, ApJ, 544, 150

Weymann, R. J. et al. 1998, ApJ, 506, 1
   
\end{chapthebibliography}

\end{document}